\documentclass[11pt]{article}
\usepackage{graphicx}
\newdimen\SaveWidth \SaveWidth=\textwidth
\newdimen\SaveHeight \SaveHeight=\textheight
\advance\SaveWidth by -\textwidth
\advance\SaveHeight by -\textheight
\divide\SaveWidth by 2
\divide\SaveHeight by 2
\advance\hoffset by \SaveWidth
\advance\voffset by \SaveHeight
\def\abs#1{\left| #1\right|}

\def\etal{{\it et al.}}
\def\abs#1{\left| #1\right|}

\def\etmiss{\slashchar{E}_T}
\def\fb{{\rm fb}}

\def\fbi{{\rm fb}^{-1}}

\let\badcite=\cite
\def\cite{~\badcite}

\def\slashchar#1{\setbox0=\hbox{$#1$}           
   \dimen0=\wd0                                 
   \setbox1=\hbox{/} \dimen1=\wd1               
   \ifdim\dimen0>\dimen1                        
      \rlap{\hbox to \dimen0{\hfil/\hfil}}      
      #1                                        
   \else                                        
      \rlap{\hbox to \dimen1{\hfil$#1$\hfil}}   
      /                                         
   \fi}                                         %

\catcode`@=11
\newdimen\vbigd@men                             

\def\vbig#1#2{{\vbigd@men=#2\divide\vbigd@men by 2%
   \hbox{$\left#1\vbox to \vbigd@men{}\right.\n@space$}}}
\catcode`@=12

\catcode`@=11
\def\citenum#1{\csname b@#1\endcsname}
\catcode`@=12

\begin{document}
\begin{titlepage}
\rightline{LBNL-54488}
\rightline{SN-ATLAS-2004-038}
\rightline{\today}
\begin{center}{\Large\bf\boldmath
Exploring  Little Higgs Models with ATLAS at the LHC}
\end{center}
\footnotetext{This work was supported in part by the Director, Office
  of Science, Office of Basic Energy Sciences, of the U.S. Department
  of Energy under  Contract No. DE-AC03-76SF00098.} 
\bigskip
\begin{center}
\bf  {G.~Azuelos$^a$,  K.~Benslama$^a$, D.~Costanzo$^b$, G.~Couture$^a$
 J.E.~Garcia$^c$,   I.~Hinchliffe$^b$,
 N.~Kanaya$^d$, M.~Lechowski$^e$, R.~Mehdiyev$^{a,f}$
G.~Polesello$^g$, E.~Ros$^c$, D.~Rousseau$^e$} 
\end{center}
\centerline{$^a${\it U. of Montreal, Montreal, Canada}}
\centerline{$^b${\it Lawrence Berkeley National Laboratory, Berkeley, CA}}
\centerline{$^c${\it IFIC, University of Valencia/CSIC, Spain}}
\centerline{$^d${ \it University of Victoria, Victoria, BC, Canada}}
\centerline{$^e${ \it University of Paris-Sud, IN2P3-CNRS, Orsay, France}}
\centerline{$^f${\it Institute of Physics, Academy of Sciences of
    Azerbaijan, Baku, Azerbaijan}}
\centerline{$^g${\it INFN, Sezione di Pavia,  Pavia, Italy}}

\bigskip
\begin{abstract}
We discuss possible searches for the new particles predicted by Little
Higgs Models at the LHC. By using a simulation of the ATLAS detector,
we demonstrate how the predicted quark, gauge bosons and additional
Higgs bosons  can be found and estimate the mass
range over which their properties can be constrained.
\end{abstract}
\end{titlepage}

\section{Introduction}

Recently, new models have been proposed as possible solutions to the
hierarchy problem of the Standard Model. This problem, also referred
to as the fine-tuning problem, is most easily phrased as follows. If
radiative corrections to the Higgs mass are computed using an
ultra-violet cut off $\Lambda$, the resulting value of the Higgs mass
is of order $\Lambda$ unless there is a very delicate
cancellation.  The ``Little Higgs'' models
\cite{Arkani-Hamed:2001ed,Arkani-Hamed:2001nc}
provide just enough  new physics to generate cancellations and
preserve a  light Higgs
boson while
raising  the ultra-violet cut off 
to a scale of several tens of TeV where constraints on new
particles from existing experiments are very weak. In the Standard
Model, there are three contributions to radiative corrections to the
Higgs mass that dominate the hierarchy problem.  In order of importance they are the top quark, the
electro-weak gauge bosons, and the Higgs itself. If the loop integrals
are cut off at scale  $\Lambda=10 $ TeV, these corrections are:
\begin{itemize}
\item from  the top loop $\delta m_h^2 =\frac{3}{8\pi^2}\lambda_t^2 \Lambda^2
\sim  (2\hbox{  TeV} )^2$;
\item from  the {$W/Z$ loops} $\delta m_h^2 \sim \alpha_w  \Lambda^2
\sim  -(750 \hbox{ GeV})^2$;
\item from  the {Higgs loop} $\delta m_h^2 \sim \frac{\lambda}{16\pi^2} \Lambda^2
\sim  -(1.25 m_h)^2$.
\end{itemize}

The  ``Little Higgs'' models implement the idea that the Higgs boson is a
pseudo-Goldstone boson \cite{Georgi:yw,Kaplan:1983sm}.
They   are constructed by embedding the  Standard
Model inside a larger group with an enlarged symmetry. The
larger symmetry is then broken at some high scale $\Lambda_H=4\pi f$,
so that the Higgs mass is protected
from radiative corrections at one loop that are 
dependent quadratically on $\Lambda_H$. From a phenomenological point of view, the
effect of this symmetry is to require  the existence of new particles
whose couplings
 ensure that the large contributions to the Higgs mass are
canceled. There must be three types of new particles corresponding to
the three contributions listed above. In this paper we  study the observability,
using the ATLAS detector at the LHC,  of these new particles. For
definiteness we will use the ``Littlest Higgs'' model
\cite{Arkani-Hamed:2002qy,Arkani-Hamed:2002qx} although many of
our results can be reinterpreted in other models. In this model,
 a new charge 2/3 quark, $T$,  which is an electroweak singlet is predicted. It
mixes with the top quark and decays via this mixing to $Wb$,  $th$
and $tZ$ final states. Since this particle is  canceling the largest
contribution to the Higgs mass, its mass cannot be too large or the
fine tuning problem will reappear: $m_T< 2 \hbox{ TeV} (\frac{m_h }{200
  \hbox{ GeV}})^2$ \cite{Arkani-Hamed:2002qy}. The model is based on an extended
gauge group  containing the gauged subgroup $[SU(2)_1\otimes U(1)_1]\otimes
[SU(2)_2\otimes U(1)_2]$ which itself implies, in addition to the $W$, $Z$ and
$\gamma$ of the Standard Model, four new gauge bosons $W_H^\pm$, $A_H$
and $Z_H$.  Since these new gauge bosons are canceling the
contributions from $W/Z$ loops, their masses  are less strongly constrained:
$M_{W_H}< 6 \hbox{ TeV} (\frac{m_h }{200 \hbox{ GeV}})^2$. 
The model also predicts new
Higgs particles forming an $SU(2)_L$ triplet ($\phi$) which contains a doubly
charged state. As the Higgs contribution to the fine-tuning
problem  is the smallest, these masses are the least constrained:  
$m_{\phi} < 10$ TeV.

 Precision
electroweak data  severely constrain the
model considered here \cite{Csaki:2002qg,Hewett:2002px}. We
will not take these constraints into account as they can be avoided in other
models whose LHC phenomenology is similar to the one we consider 
\cite{Chang:2003zn,Kaplan:2003uc}.

The rest of this paper is organized as follows. We discuss the
channels available for the discovery and measurement of the properties
of the $T$ quark in the next section. We then discuss the new gauge
bosons and the doubly charged Higgs boson and finally we draw
some conclusions regarding the LHC sensitivity to Little Higgs
models.  Specific masses are chosen for the the simulations shown below. Our
main purpose is to assess the potential of ATLAS for the Little Higgs
Model, so we have chosen  masses that are  rather large as observation is more
difficult in these cases due to the smaller event rates.  In a few
cases where we were concerned that the signal to background ratio
might get worse with reduced masses, we show a lower mass case to
allay this concern. A Higgs mass of
120 GeV was assumed and we comment on the effect of this choice in the
conclusion. We further assume that the Higgs boson has been discovered and
that its mass is known.

PYTHIA 6.203 \cite{Sjostrand:2001yu} with suitably
normalized
 rates was used to generate
events which were passed through the ATLAS fast
simulation\cite{atlfast} which provides a parametrized response of the
ATLAS detector to jets, electrons, muons, isolated photons and missing
transverse energy. This fast simulation
 has been validated using a large number of studies \cite{tdr} where it was
compared with, and adjusted to agree with, 
 the results of a full, GEANT based, simulation
\cite{atlsim}. The performance for high luminosity environment is
assumed. The fast simulation provides a standard definition
of isolated leptons and photons that is used throughout. Using these
definitions, fake electrons arising from misidentified jets are
negligible for our purposes. Jets are
reconstructed using a cone algorithm with a cone of size $\Delta
R=0.4$. Performance for the high luminosity ($10^{34}$ cm$^{-2}$
sec$^{-1}$ is assumed. 
There is one case, that of identifying jets containing
$b-$hadrons ($b-$tagging), where the validation was in a kinematic
range different from that needed in this study. In this case, a full
simulation was performed and its results used to reparametrize the
tagging efficiency and the rejection against non$-b$ jets used in 
fast simulation. (See details in Section~3.3.) 
 The event selections  are based on the characteristics of the signal
 being searched for, and are such that they will pass
the ATLAS trigger criteria.  The most important triggers arise   from
the isolated leptons, jets  or
photons present in the signal. These event selections exploit the
experience gained in devising the stratgies for other new physics
searches \cite{tdr}. The  selections have not been optimized in
detail for the particular cases under study. In some cases, the cuts
have been  varied  when the particle masses were changed; for
example raising the threshold on  particular physics objects as the
mass of the new particle increases. It is not  claimed that the event
selections are optimal. Detailed optimization is not wise at this
stage due to uncertainties in the background estimates.

PYTHIA was also  use for simulation of
the backgrounds. In a few cases, discussed explicitly below, other event
 generators were used if the backgrounds are needed in regions of
 phase space where PYTHIA is known to be less reliable (mainly for
 processes with large numbers of well-separated jets).
Systematic uncertainties in the level of the
backgrounds are, in many cases, difficult to estimate. Since we are only
interested in the observability of the signals, but are not, at this
stage, attempting to evaluate the precision with which cross-sections
can be measured, these uncertainties are not expected to affect
significantly the results. Indeed, in most cases, the signals appear as
clear peaks above a smooth background. The precision with 
 which masses can be measured will ultimately depend on the
 calibration of the detector. Previous studies can be consulted for a
 discussion of these issues \cite{tdr}.

\section{Search for $T$ and determination of its properties}

\begin{figure}
\centerline{\includegraphics[width=4in]{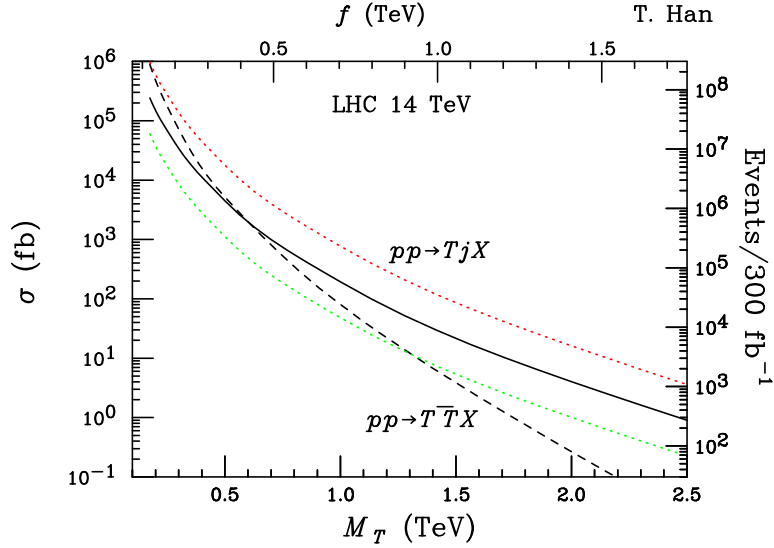}}
\caption{Figure showing the production rate of the $T$ quark at the
  LHC as a function of its mass \cite{Han:2003wu}. The heavy dashed
  line shows the pair production and the solid and two dotted lines
  the single production rate for three value of $\lambda_2/\lambda_1$; 
  from highest to lowest $\lambda_2/\lambda_1=2,1,0,5$. (We are
  grateful to T.~Han for providing this figure.)
 \label{han1}} 
\end{figure}

The $T$ quark can be produced at the LHC via two mechanisms: 
QCD production via the processes $gg\to T\overline{T}$ and $q\overline{q}\to
T\overline{T} $ which depend only on the mass of $T$; and production via 
$W$ exchange  $q b  \to q' T$ which leads to a single $T$ in the final state
and therefore falls off much more slowly as $m_T$ increases.  This
latter process depends on the model parameters and, in particular, upon
the mixing of the $T$ with the conventional top quark. 
The Yukawa couplings of the new  $T$ are given by two constants $\lambda_1$
and $\lambda_2$
$$ \lambda_1(i{Qht_r} +f {T_Lt_r} -\frac{1}{2f}{T_Lt_rhh^\dagger})
+ \lambda_2f({T_LT_R})$$
where $h$ is the Standard Model Higgs doublet, $Q$ is a doublet
containing the left-handed top and bottom quarks ($t_L$, $b_L$), and
$t_R$ is the right-handed top quark; (for details see \cite{Han:2003wu} whose notation is
followed here). The physical top quark mass eigenstate is a mixture of
$t$ and $T$. 
These
couplings contain three parameters $\lambda_1$, $\lambda_2$ and $f$
that determine the masses of $T$ and the top quark as well as  their
mixings. Two of the parameters  can be reinterpreted as the top mass and the $T$
mass. The third
can then be taken to be $\lambda_1/\lambda_2$. This  determines the
mixings and hence the coupling strength $TbW$ which controls the
production rate via the $q b  \to q' T$ process. The production rates
are shown in Figure \ref{han1} from \cite{Han:2003wu}. It can be seen
that  single
production dominates for masses above 700 GeV. As we expect that we
are sensitive to masses larger than this, we consider only the single production
process in what follows. We assume the following cross-sections: for
$m_T=0.7,1.0,1.3,1.6$ TeV, $\sigma=1000,200,45,12$ $\fb$ for
$\lambda_1/\lambda_2=1$. Events  generated using PYTHIA were
normalized to these values.

The decay rates of $T$ are as follows
$$ \Gamma(T \to tZ) = 
{1\over 2} \Gamma(T \to bW)=\Gamma(T \to th) ={\kappa^2\over 32\pi}M_T$$
with $\kappa={\lambda_1^2/\sqrt{\lambda_1^2+\lambda_2^2}}$ implying that
$T$ is  a narrow resonance.
The last of these decays  would be expected for a charged
2/3 4$^{th}$ generation quark; the first two are special to the
``Little Higgs Model''. We now discuss the reconstruction in these
 channels.

\subsection{$T\to Zt$}

This channel can be observed via the final state $Zt \to \ell^+ \ell^-
\ell \nu b$, which implies that the events contain three isolated leptons, a pair of which
reconstructs to the $Z$ mass,  one $b-$jet and
missing transverse energy. The background is dominated by $WZ$, $ZZ$
and 
$tbZ$. The last  cannot be simulated using PYTHIA, and was therefore 
generated using CompHep \cite{Pukhov:1999gg}. 
 Events were selected as follows.
\begin{itemize}
\item Three isolated leptons (either $e$ or $\mu$) with  $p_T>$ 40
  GeV and $\abs{\eta}<2.5$. One of these is required to have $p_T>$ 100
  GeV. 
\item No other leptons with $p_T>$ 15
  GeV. 
\item $\etmiss > 100$ GeV.
\item At least one tagged $b-$jet  with $p_T>$ 30
  GeV.
\end{itemize}
The presence of the leptons ensures that the events are triggered.
A pair of leptons of same flavor and opposite sign is required to have
an invariant mass within 10 GeV of  $Z$ mass. The efficiency of 
these cuts is 3.3\% for $m_T=1000$ GeV.  The third lepton is
then assumed to arise from a $W$ and the $W$'s momentum reconstructed
 using it and the measured $\etmiss$. 

\begin{figure}
\centerline{\includegraphics[width=4in]{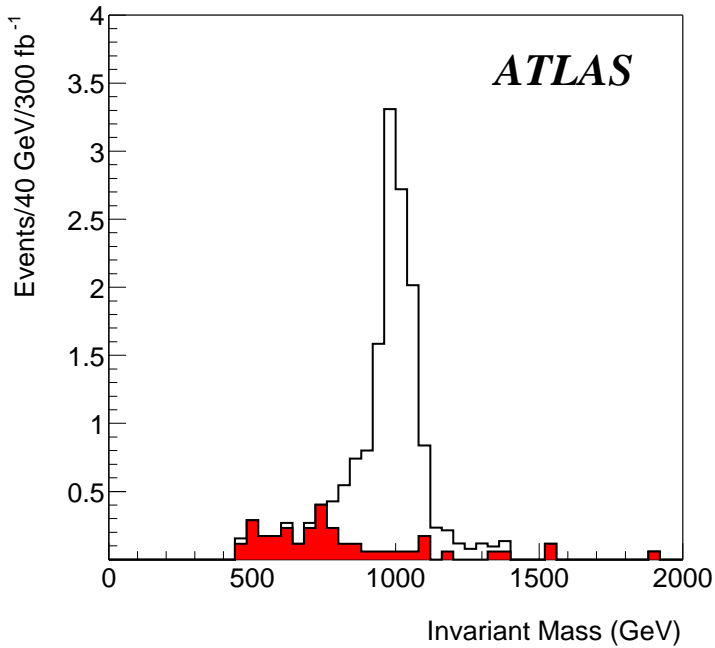}}
\caption{Reconstructed mass of the $Z$ and $t$ (inferred from the
  measured lepton,  $\etmiss$, and tagged $b-$jet). The signal $T \to Zt$  is
  shown for a mass of 1000  GeV. The background, shown as the filled histogram, is
  dominated by $WZ$ and $tbZ$ (the latter is larger)   production. 
The signal event rates
  correspond to $\lambda_1/\lambda_2=1$ and a $BR(T\to ht)$ of
  25\%. More details can be found in Ref~\cite{costanzo}.
 \label{cost1}}
\end{figure}

The invariant mass of the $Zt$
system can then be reconstructed by including the $b-$jet. 
This is shown in Figure~\ref{cost1}
for $m_T=1000$ GeV where a clear peak is visible above the
background. Following the cuts, the background is dominated by $tbZ$
which is more than 10 times greater than all the others combined. The
cuts accept 0.8\% of this background \cite{costanzo}.

Using this analysis, the discovery
potential in this channel can be estimated. The signal to background
ratio is excellent as can be seen from Figure~\ref{cost1}. Requiring a
peak of 
at least  $5\sigma$ significance containing at least 10 reconstructed events implies that
for
$\lambda_1/\lambda_2=1(2) $
and 300 fb$^{-1}$ the quark of mass $M_T<1050 (1400) $ GeV is observable.
At these   values, the single $T$ production process dominates,
justifying {\it a posteriori} the neglect of $T\overline{T}$
production in this simulation. 

\subsection{$T\to Wb $}

\begin{figure}
\centerline{\includegraphics[width=4in]{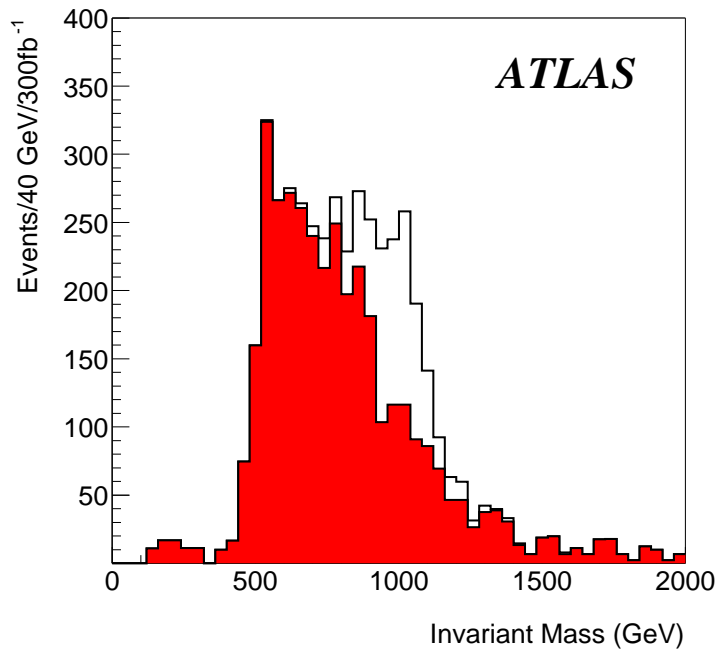}}
\caption{Reconstructed mass of the  $W$ (inferred from the
  measured lepton   and $\etmiss$) and tagged $b-$jet. The signal arises from the decay $T \to Wb$ and is
  shown a for mass of 1000  GeV. The background,  shown separately as the
  filled histogram, is
  dominated by $t\overline{t}$ and single top production (the former is larger). The signal event rates
  correspond to $\lambda_1/\lambda_2=1$ and a $BR(T\to Wb)$ of 50\%. More details can be found in Ref~\cite{costanzo}.
 \label{cost2}}
\end{figure}

This channel can be reconstructed via the final state $\ell\nu b$. The
following event selection was applied.
\begin{itemize}
\item At least one  charged  lepton with $p_T>$100 GeV.
\item One $b$-jet with $p_T>$ 200 GeV.
\item No more than 2 jets with $p_T>30 $ GeV.
\item Mass of the pair of jets with the highest $p_T$  is greater than $200$ GeV.
\item $\etmiss>$100 GeV.
\end{itemize}
The lepton provides a trigger. 
The efficiency of this selection for a $T$ of mass 1 TeV is 14\%. The
backgrounds arise from $t\overline{t}$, single top production and QCD
production of $Wb\overline{b}$.  These are estimated using PYTHIA for
the first one, CompHep \cite{Pukhov:1999gg} for the second  and AcerMC
\cite{Kersevan:2002dd} for the last.
 The requirement of only one tagged
$b-$jet and the high $p_T$ lepton are effective against all of these
backgrounds. The requirement of only two energetic jets  is powerful against the dangerous
$t\overline{t}$ source where the candidate $b-$jet arises from the $t$ and
the lepton from the $\overline{t}$. These cuts  reduce the
total $t\overline{t}$ and $Wb\overline{b}$ by factors of $2.5\times
10^{-5}$ and $7.5\times
10^{-5}$ respectively. 
Figure~\ref{cost2} shows the reconstructed mass of the $Wb$ system
where the $W$ momentum is inferred from the
  measured lepton  $\etmiss$ using the $W$ mass as a constraint. 
The plot shows the signal arising from $T$ of mass 1 TeV as a
peak over the remaining background. The signal to background ratio is
somewhat worse than in the previous case primarily due to the
$t\overline{t}$ contribution.

From this analysis, the discovery
potential in this channel can be estimated. For
$\lambda_1/\lambda_2=1(2) $
and 300 fb$^{-1}$ $M_T<2000 (2500) $ GeV has at least a $5\sigma$
significance.

\subsection{$T\to ht$}
\label{sec:ht}

\begin{figure}
\centerline{\includegraphics[width=4in]{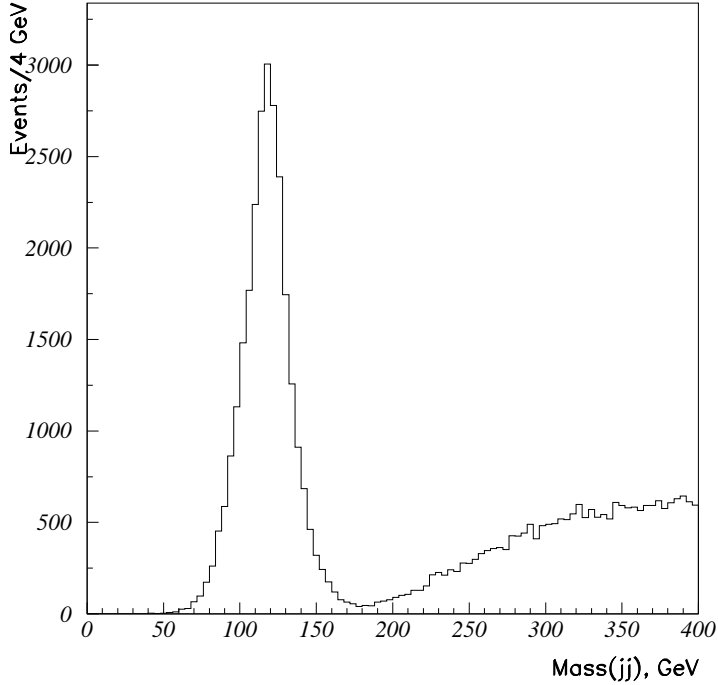}}
\caption{Plot showing the dijet mass distribution arising from 
  the decay $T \to ht$. All combinations of jets are shown. 
 \label{rashid1}}
\end{figure}

\begin{figure}
\centerline{\includegraphics[width=4in]{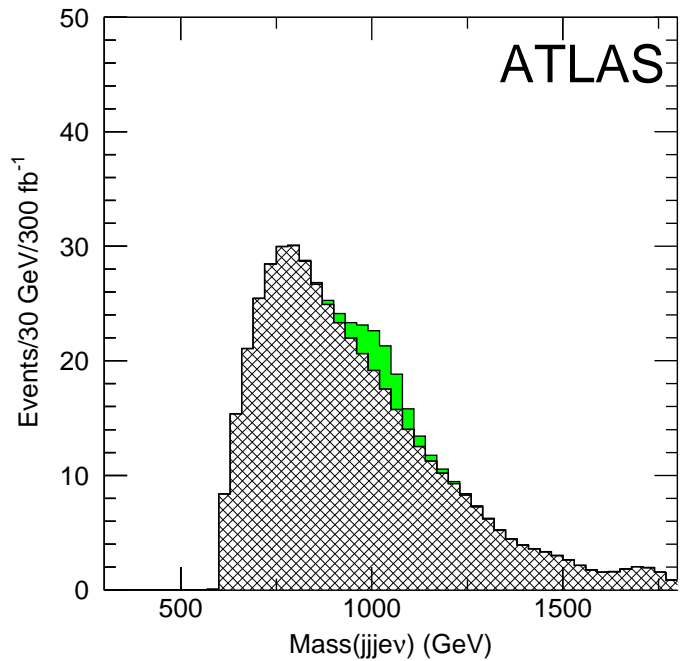}}
\caption{Reconstructed mass of the $W$ (inferred from the isolated
  lepton and missing transverse energy) and three jets, two of which
  are required to have an invariant mass consistent with the Higgs
  mass. The signal arises from the decay $T \to ht$ and is
  shown for a mass of  1000 GeV. The background, shown in cross-hatching, is
  dominated by $t\overline{t}$ production. The signal event rates
  correspond to $\lambda_1/\lambda_2=1$ and a $BR(T\to ht)$ of 25\%.
  \label{rashid2}}
\end{figure}

In this final state, the event topology depends on the 
Higgs mass. For a Higgs mass of 120 GeV the decay to $b\overline{b}$
dominates. The semileptonic top decay $t \to Wb \to \ell\nu b$
produces a lepton that can provide a trigger. The final state containing
of an isolated lepton and several jets then needs to be
identified.
    The initial event selection is as follows.
\begin{itemize}
\item One isolated $e$ or $\mu$ with $p_T>$  100 GeV and $\abs{\eta}<2.5$.
\item Three jets  with $p_T>$  130 GeV.
\item At least one jet tagged as a $b-$jet.
\end{itemize}
 The
dijet mass distribution of all pairs of  jets in events from $T$
production that pass these
cuts is shown in
Figure~\ref{rashid1}. A  clear peak at the Higgs mass is visible. It
should be noted that the jets in this plot are not required to be tagged
as $b-$jets. The requirement of more than one jet tagged as a
$b-$jet lowers the efficiency and is not necessary to extract a
signal.
Events
were further selected by requiring that at least one di-jet
combination have a mass in the range 110 to 130 GeV. If there is a
pair of jets with invariant mass in the range 70 to 90 GeV, the event
is rejected; this will help to reduce the $t\overline{t}$ background.
 The measured
missing transverse energy and the lepton were then combined using the
assumption that they arise from a $W \to \ell\nu$ decay. Events that
are consistent with this are retained and the $W$ momentum inferred. The
invariant mass of the reconstructed $W$, $h$ and one more jet is
formed and the result shown in Figure~\ref{rashid2}. For a  mass of 1000 
GeV, the cuts accept 2.3\% of the signal events. The width of the reconstructed $T$ resonance is dominated by
experimental resolution.

The signal shown in Figure~\ref{rashid2} assumes that
$\lambda_1/\lambda_2=1$. The background is
dominated by $t\overline{t}$ events as a semileptonic decay of either
$t$ or $\overline{t}$ produces all objects necessary to pass the event
selection. Only the different kinematics in the signal and background
can distinguish them.
For example, the lighter top quark implies that the lepton has a
softer $p_T$ spectrum than that from the signal. 
  The larger background implies that discovery in this mode is more
  difficult than the cases discussed above. However once the $T$ has
  been discovered in another channel, the peak shown  in
  Figure~\ref{rashid2}, which has a significance above
$4\sigma$ that is sufficient to confirm  a signal 
and constrain the branching ratio.  

\begin{figure}
\centerline{\includegraphics[width=4in]{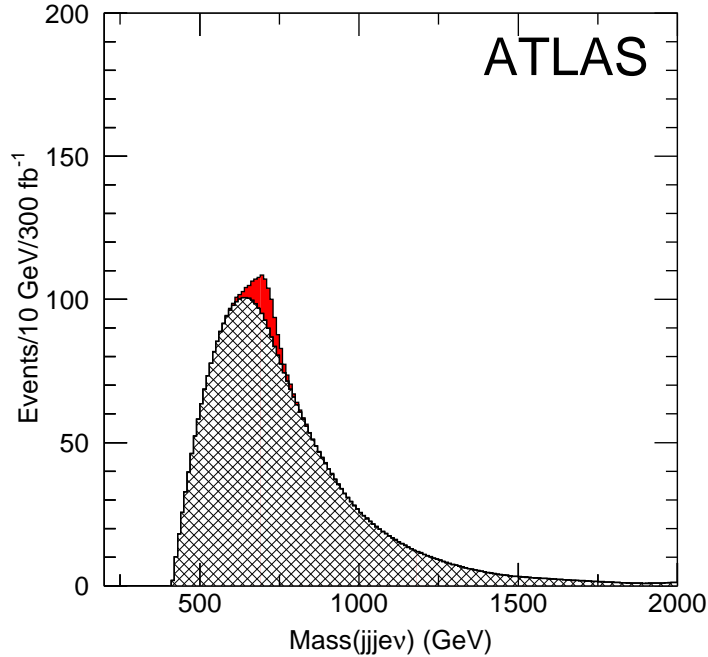}}
\caption{As for Figure~5 except that the $T$ mass is 700 GeV and
  the cuts are looser (see text).
  \label{rashid3}}
\end{figure}

As the mass is reduced towards the top mass, the signal
becomes more difficult to extract from  the  $t\overline{t}$ background as the leptons
and jets from the $T$ decay become softer. In order to investigate
this posssible difficulty,   a $T$ mass of 700 GeV was simulated. The cuts on the
jets and leptons must be relaxed,  to 90 GeV for the jets and
to  70 GeV for the lepton. With these values, the signal efficiency is only
reduced to 1.1\%.
 Figure~\ref{rashid3} shows the resulting distribution. 
The larger background results in a less significant
signal.  The signal shown is approximately $3\sigma$ significance
which will not provide a discovery, but would confirm a signal seen in
another channel and will enable a constraint on the couplings to be
deduced. Of course, a larger value of $\lambda_1/\lambda_2$ will result
in a clearer signal.

\section{Search for new gauge bosons}

The model predicts the existence of one charged $W_H$ and two neutral
($Z_H$ and $A_H$) heavy
gauge bosons.  $W_H$ and $Z_H$ are almost  degenerate in mass and are
typically 
heavier than $A_H$. All these bosons are likely to be discovered via their decays to
leptons. However, in order to distinguish these gauge bosons from those
that can arise in other models, the characteristic decays $Z_H\to Zh$
and $W_H\to Wh$ must be observed \cite{Burdman:2002ns}. The properties
of the new gauge bosons are determined by the couplings of the gauge
theory  $[SU(2)_1\otimes U(1)_1]\times [SU(2)_2\otimes U(1)_2]$
implying
 two couplings in addition to those of  the Standard Model
$[SU(2)_L\otimes U(1)]$ group. These additional parameters can be 
taken  to be two angles $\theta$ and
$\theta^{\prime}$ (somewhat analogous to $\theta_W$ of the Standard
Model). Once the masses of the new bosons are specified,
$\theta$ 
determines the couplings of  $Z_H$ and $\theta^{\prime}$ those of $A_H$
(if it is assumed that there are no anomalies). In the case of $Z_H$, the branching ratio
into $e^+e^-$ and $\mu^+\mu^-$ rises with $\cot\theta$ to an asymptotic
value of 4\%.

\subsection{Discovery of $Z_H$ and $A_H$}

\begin{figure}
\centerline{\includegraphics[width=4in]{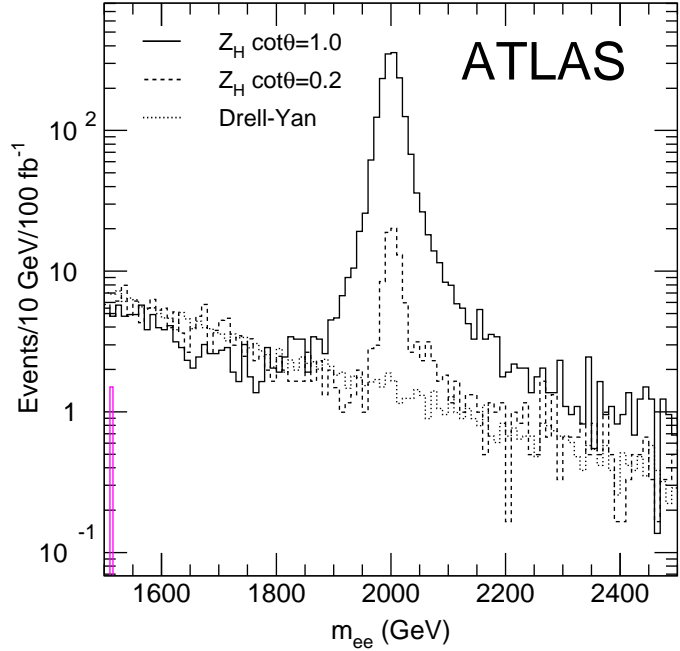}}
\caption{Plot showing the 
  $e^+e^-$ mass distribution arising from a $Z_H$ of mass of 2 TeV for
  $\cot\theta =1$ (upper, solid, histogram) and $\cot\theta =0.2$
  (middle, dashed,
  histogram). The  lowest, dotted. histogram 
  shows the distribution from background only.
 \label{zpeak}}
\end{figure}

\begin{figure}
\centerline{\includegraphics[width=4in]{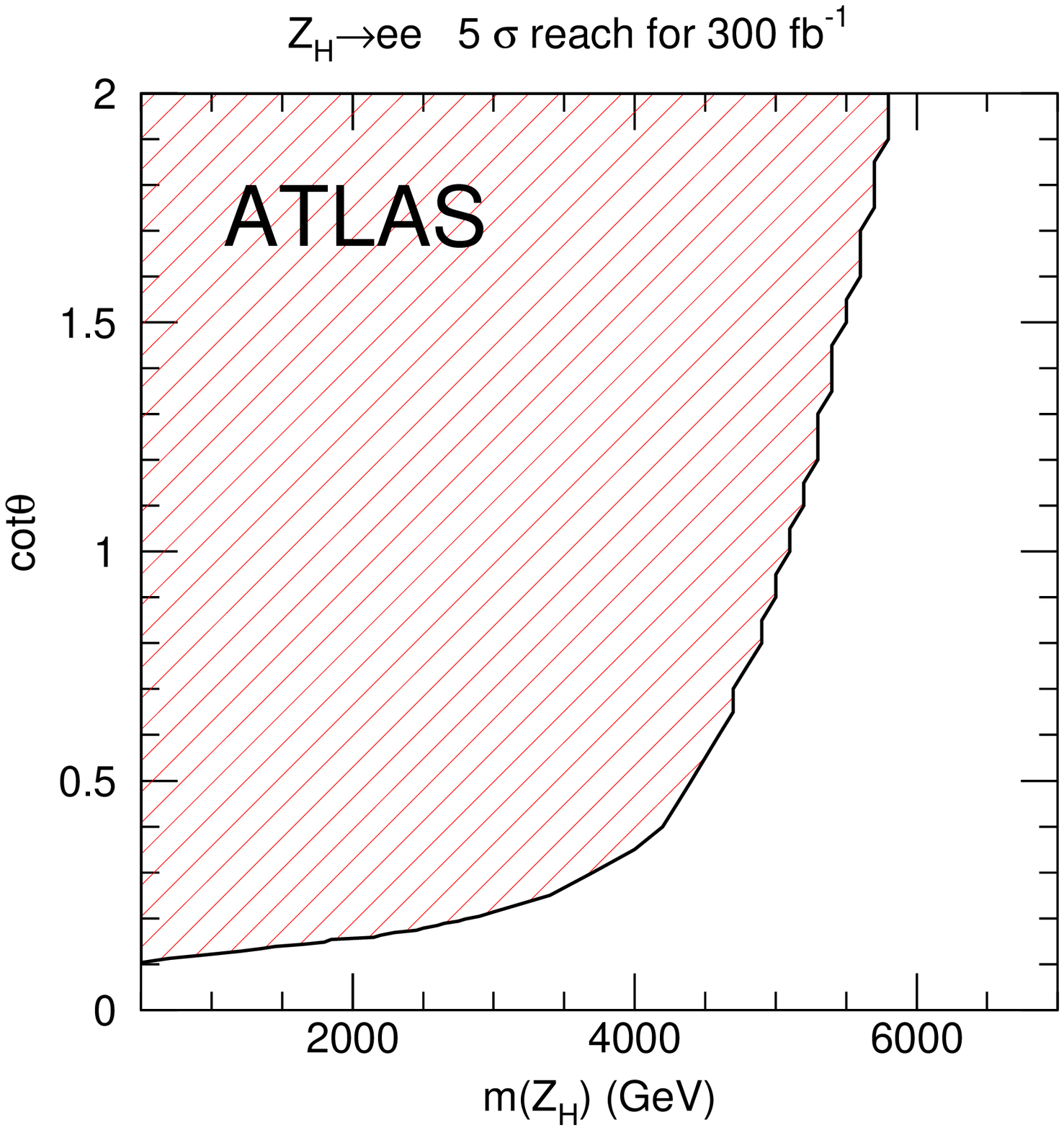}}
\caption{Plot showing the accessible region (shaded) in the channel $Z_H \to
  e^+e^-$  as a function of the mass and the mixing
  $\cot\theta^{\prime}$. 
 \label{reachz}}
\end{figure}

\begin{figure}
\centerline{\includegraphics[width=4in]{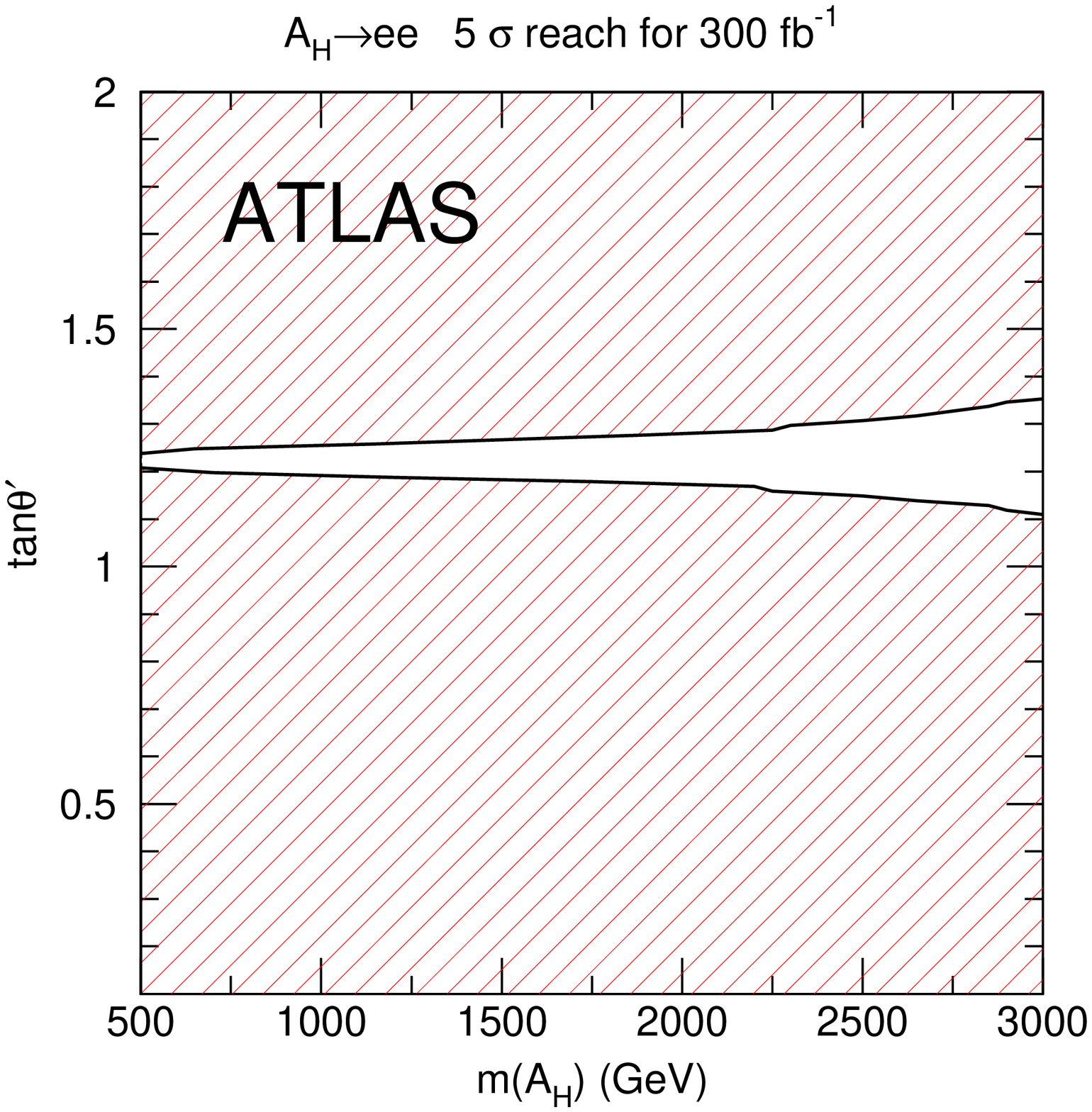}}
\caption{Plot showing the accessible region (shaded) in the channel $A_H \to
  e^+e^-$  as a function of the mass and the mixing
  $\tan\theta^{\prime}$. 
  significance.
 \label{reacha}}
\end{figure}

 A search for a peak in the invariant mass distribution of either  $e^+e^-$ or $\mu^+\mu^-$
is sensitive to the presence of $A_H$ or $Z_H$. As an example,
Figure~\ref{zpeak} shows the  $e^+e^-$ mass distribution arising from a $Z_H$ of mass of 2 TeV for
  $\cot\theta =1$ and $\cot\theta =0.2$. The production cross-section
  for the former (latter)  case is 1.2  (0.05)  pb \cite{Han:2003wu}. Events were required to have
  an isolated $e^+$ and $e^-$ of $p_T>20$ GeV and  $\abs{\eta}<2.5$ which
  provides a trigger. The
  Standard Model background shown on the plot arises from the
  Drell-Yan process. In order to establish a signal we require at
  least 10 events in the peak of at least $5\sigma$
  significance. Figure~\ref{reachz} shows the accessible region as a
  function of  $\cot\theta$ and $M_{Z_H}$; a slightly greater reach can
  be obtained by including the  $\mu^+\mu^-$ channel. Except for very
  small values of  $\cot\theta$, where the leptonic branching ratio is
  very small, the reach covers the entire region expected
  in the model. A similar search for $A_H$ can be carried out and the accessible region as a
  function of  $\tan\theta^{\prime}$ and $M_{A_H}$ is shown in
  Figure~\ref{reacha}. Masses greater than 3 TeV are not shown as
  these are not allowed in the model. There is a small region 
around  $\tan\theta^{\prime}\sim 1.3$ where the branching ratio to
$\mu^+\mu^-$ and  $e^+e^-$ is very small and the channel is
insensitive.

\subsection{Discovery of $W^{\pm}_H$ }

\begin{figure}
\centerline{\includegraphics[width=4in]{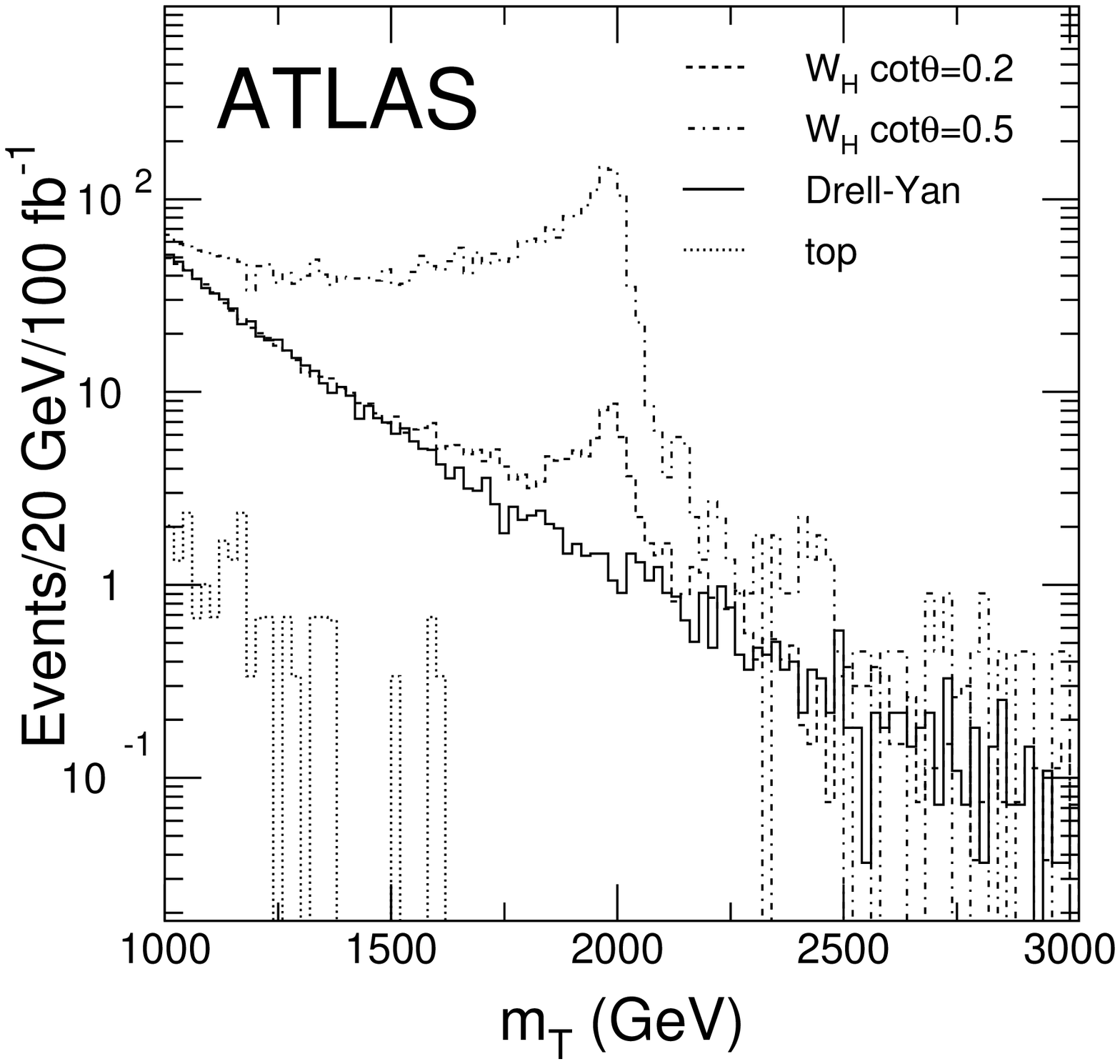}}
\caption{Plot showing the 
  transverse  mass distribution from $\etmiss$ and $e$ arising from a $W_H$ of mass  2 TeV for
  $\cot\theta =0.5$ (upper histogram) and $\cot\theta =0.2$ (second
  highest histogram). The third highest histogram shows
  shows the distribution from the Drell-Yan background. The very small
  background from $t\overline{t}$ is also shown.
 \label{wpeak}}
\end{figure}

\begin{figure}
\centerline{\includegraphics[width=4in]{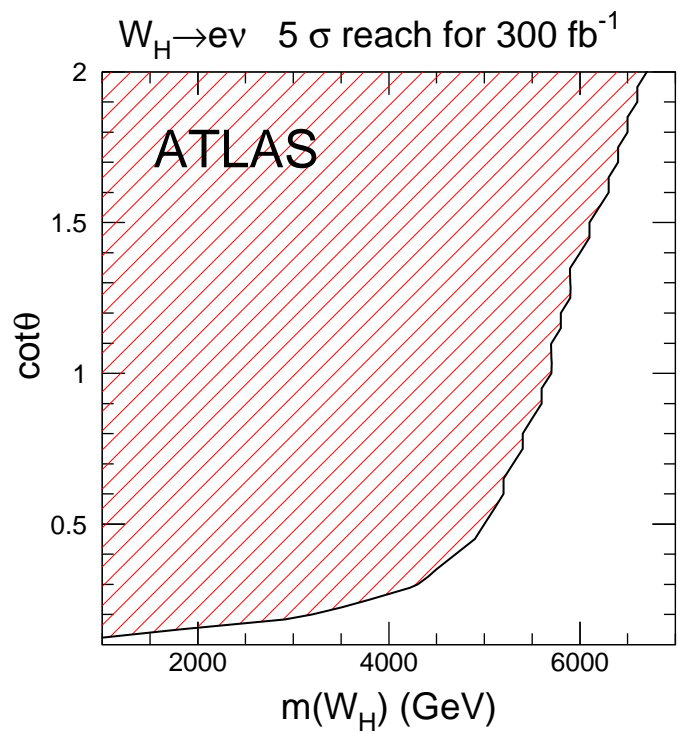}}
\caption{Plot showing the accessible region (shaded) in the channel $W_H \to
  \ell\nu$  as a function of the mass and the mixing
  $\cot\theta$.
 \label{reachw}}
\end{figure}

The decay $W^{\pm}_H\to \ell\nu$ manifests itself via events that
contain an isolated charged lepton and missing transverse
energy. Events were selected by requiring an isolated electron with
$e^-$ or $e^+$ of $p_T>200$ GeV, $\abs{\eta}<2.5$ and $\etmiss> 200$
GeV.
 The transverse mass from
$\etmiss$  and the observed lepton is formed and the signal appears
as a peak in the distribution as illustrated in Figure~\ref{wpeak}. There is a small
background from $t\overline{t}$ events as can be seen from the plot; the
main  background arises from  $\ell\nu$ production via a virtual
$W$, labeled as Drell-Yan on the figure. In order to establish a signal we require at
  least 10 events in the signal region  of at least $5\sigma$
  significance.
  Figure~\ref{reachw} shows the accessible region as a
  function of  $\cot\theta$ and $M_{W_H}$; a slightly greater reach can
  be obtained by including the  $\mu\nu$ channel. Except for very
  small values of  $\cot\theta$, the reach covers the region allowed
  in the model.

\subsection{Measurement of $Z_H\to Zh$,  $A_H\to Zh$ and $W_H\to Wh$}

\begin{figure}
\centerline{\includegraphics[width=4in]{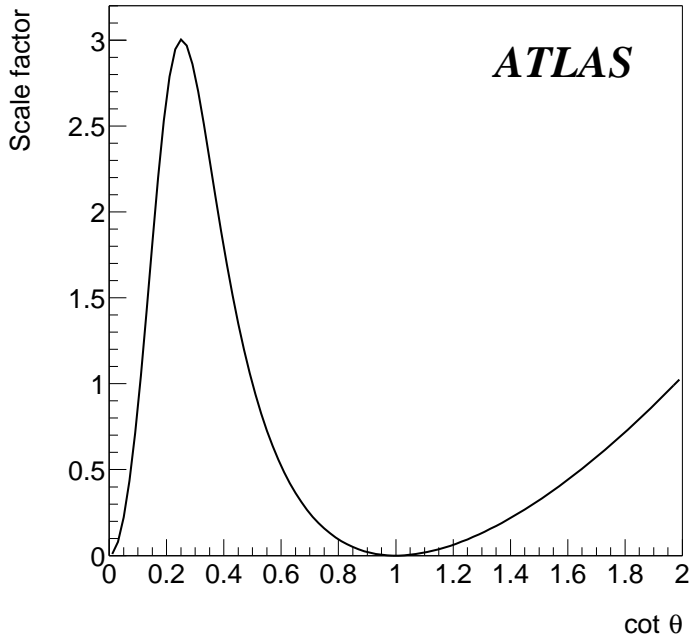}}
\caption{The $\cot\theta$ dependence of the
production rate times branching ratio $Z_H \to Zh$.
 \label{azhbr}}
\end{figure}

Observation of the  cascade decays $Z_H\to Zh$, $A_H\to Zh$, and
$W_H\to Wh$ provides  crucial evidence that  an observed new gauge boson is
of the type predicted in the Little Higgs Models. We begin by
discussing $Z_H\to
Zh$ \cite{jose}. For our choice of Higgs mass,  the decay $Zh \to \ell^+\ell^-
b\overline{b}$ results in a final state with two $b-$jets that
reconstruct to the Higgs mass and a  $\ell^+\ell^-$ pair that
reconstructs to the $Z$ mass.  The coupling $Z_HZh$ is proportional to
$\cot 2 \theta$. When combined with the coupling  of $Z_H$ to quarks
that controls the production cross-section, the $\cot\theta$
dependence of the rate in this channel is shown in Figure~\ref{azhbr}, which
illustrates the difficulty of observation when $\cot\theta \sim 1$.

\begin{figure}
\centerline{\includegraphics[width=4in]{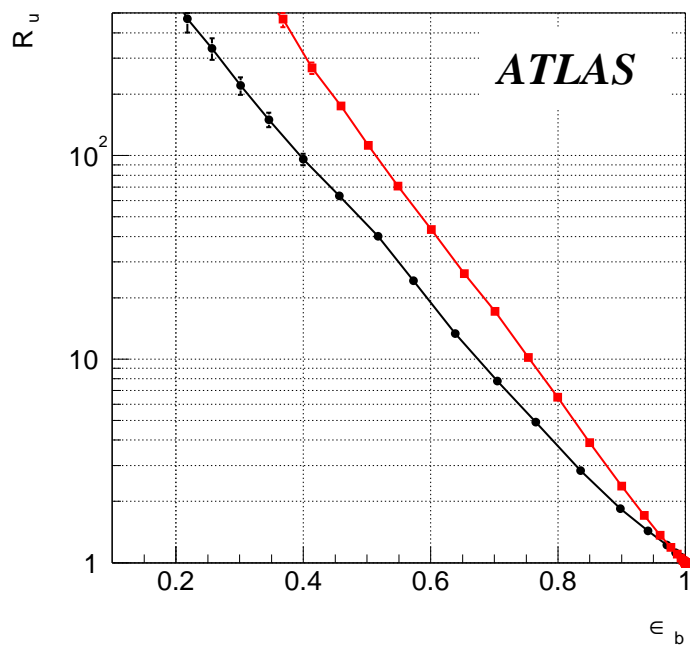}}
\caption{ Plot showing the tagging efficiency for $b-$jets as a
  function of the rejection factor against light quark jets. The upper
  curve shows the result from the benchmark ATLAS sample of bottom
  quarks from a Higgs decay  of mass 400 GeV produced in
  association with a $W$ \cite{tdr}. The lower
  curve shows the result from the higher energy $b-$quarks from the
  $Z_H\to Zh$ sample.
 \label{btag}}
\end{figure}

The transverse momentum of the $b-$jets
from this Higgs decay is of order $0.25M_{Z_H}$ since the Higgs is
boosted.  This is larger than
the $p_T$ of the jets in the 
full simulation samples  of $Wh$ events \cite{tdr} that were used for parameterizing the fast
simulation.  The average transverse momentum of the $h$ is of order
its mass in the $Wh$ sample.  Therefore a full, GEANT based
\cite{atlsim},  
simulation
was run and the events reconstructed to check the $b-$tagging
assumptions in the fast simulation. 
 A sample of events arising from $Z_H$ of mass 2 TeV  decaying to $Zh$
 was
produced. The decay of $h$ was forced to either $b\overline{b}$ or
$u\overline{u}$ so as to provide a sample of $b-$ and light quark jets
in the same kinematic region.
Figure~\ref{btag} shows the efficiency for tagging the $b-$jets as a
function of the rejection that the same algorithm obtains for
$u-$quark jets in these samples.  For comparison, the result is also  shown for a sample
from $Wh$ production with a Higgs mass of 400 GeV. This latter sample
was one of the ones used to obtain the parametrized response for the
fast simulation. A comparison of the two curves shows a 
degradation relative to the existing benchmark. Nevertheless, the
rejection against light quark jets is still more than adequate. 
For the plots in the
remainder of this section, we have reparametrized the fast simulation
to take account of this degradation. We use a tagging efficiency of
40\% and the corresonding rejection of a factor of 100 against light
quark jets.

Following this aside, we now extract a signal from the  $Z_H\to Zh$
 state using the following event selection.
 \begin{itemize}
 \item Two leptons of opposite charge and same flavor with $p_T>6(5) $
   GeV for muons (electrons) and $\abs{\eta}<2.5$.  One of them is
   required to satisfy $p_T>25 $
   GeV in order to provide a trigger.
\item  The lepton  pair has a mass between 76 and 106 GeV
\item Two reconstructed $b-$jets with $p_T>25 $ GeV
   and $\abs{\eta}<2.5$, which are within  \\ $\Delta R =\sqrt{(\Delta
     \eta)^2+(\Delta \phi)^2} < 1.5$.
\item The $b-$jet pair should have a mass between 60 and 180 GeV.
 \end{itemize}

\begin{figure}
\centerline{\includegraphics[width=4in]{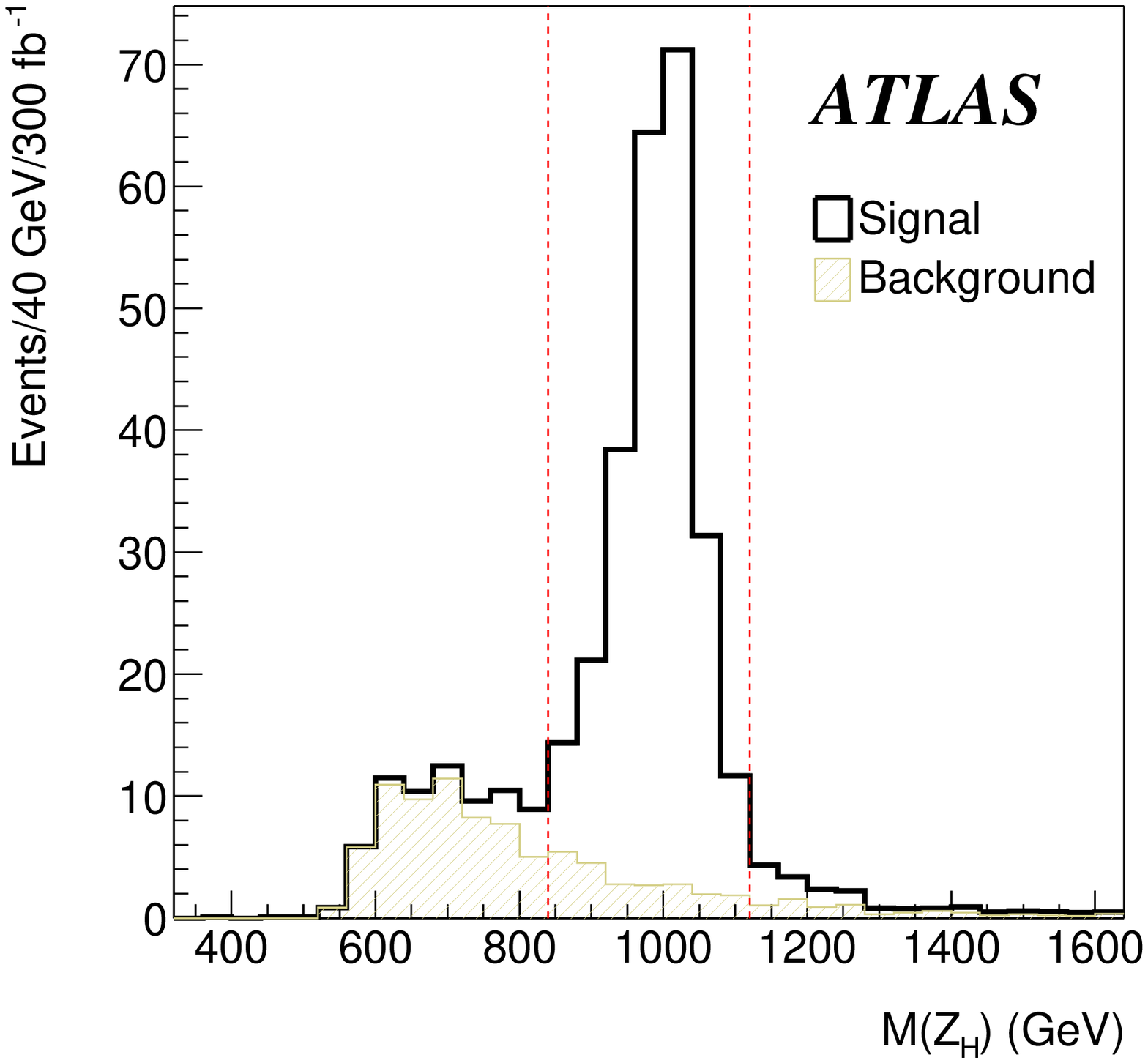}}
\caption{Invariant mass of the $Zh$ system reconstructed from the
  $\ell^+\ell^- b\overline{b}$ final state showing the signal from a $Z_H$
  of mass 1000 GeV with $\cot\theta=0.5$ above the Standard Model
  background. The vertical lines define the signal region.
 \label{azh1}}
\end{figure}

\begin{figure}
\centerline{\includegraphics[width=4in]{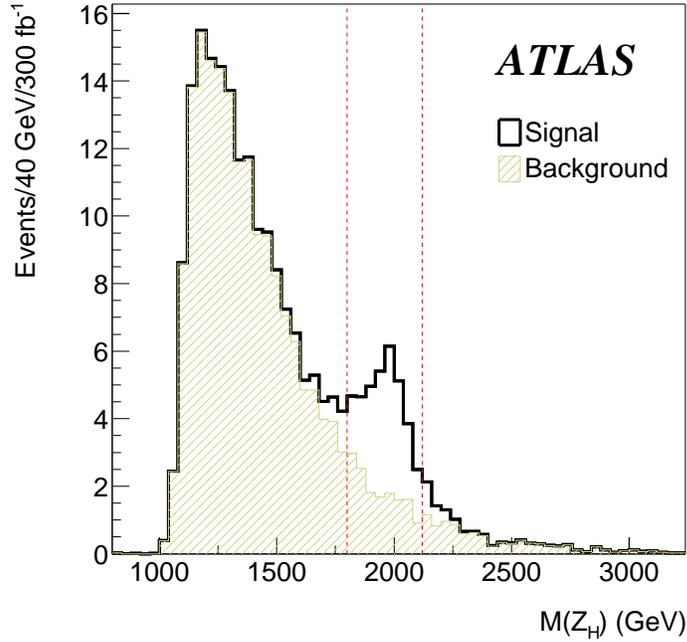}}
\caption{As in Figure~\ref{azh1} except that the $Z_H$ mass is  2 TeV.
 \label{azh2}}
\end{figure}

These cuts accept 35\% of the $\ell\ell b\overline{b}$ events.
 The mass of the reconstructed $Zh$ system is shown in
Figures~\ref{azh1} and \ref{azh2} for $Z_H$ masses of 1000 and 2000
GeV and $\cot\theta=0.5$.   In the case of  Figure~\ref{azh2}, the mass of $Z_H$ is so large  that the
two jets from the Higgs decay coalesce into a single jet. In this
case, therefore, 
 only one  $b-$jet was required with $p_T>500$ GeV and the invariant
mass of that jet  required to be in the Higgs window.
The presence of a leptonic $Z$ decay in the signal ensures that the background arises primarily from $Z+jet$
final states. Approximately $10^{-5}$ of the $Z+jet$ events pass the
kinematic cuts. The signal shown in Figure~\ref{azh2} corresponds to
$S/\sqrt{B}=5$ in  a window of width 300 GeV around the peak. There is
an uncertainty in the  estimate of the
background which was generated using PYTHIA.
 However since a peak is clearly
visible, the background can be constrained from the sidebands once
data exists.

\begin{figure}
\centerline{\includegraphics[width=4in]{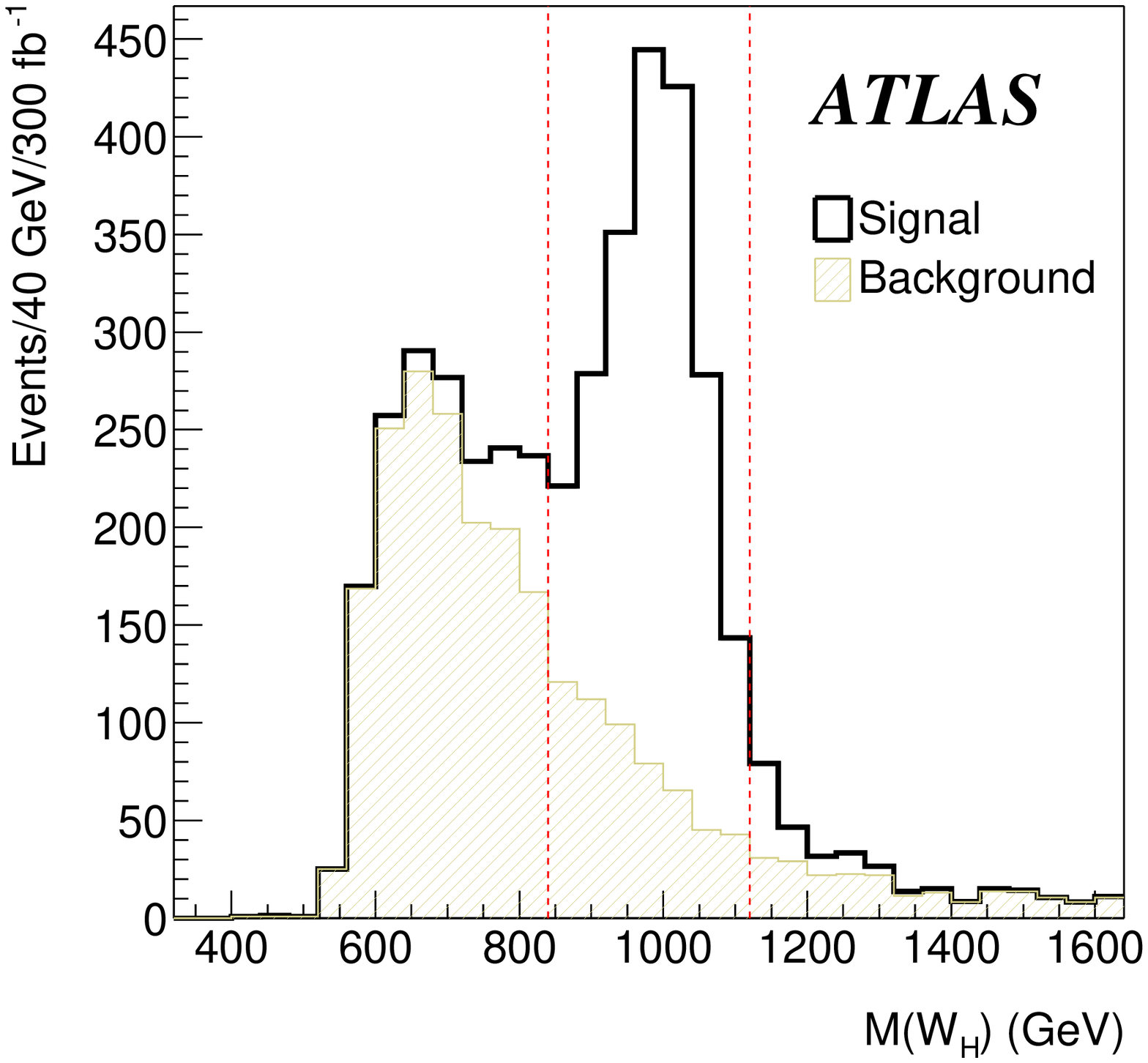}}
\caption{Invariant mass of the $Wh$ system reconstructed from the
  $\ell^+\nu b\overline{b}$ final state showing the signal from a $W_H$
  of mass 1000 GeV with $\cot\theta=0.5$ above the Standard Model
  background. The vertical lines define the signal region.
 \label{wh2}}
\end{figure}

A similar method can be used to reconstruct the $W_H\to Wh\to \ell\nu
b\overline{b}$ decay \cite{jose}. The $b-$jet selections were the same as above
while the lepton selection is now as follows.
\begin{itemize}
\item  One isolated $e$ or $\mu$ with $p_T>$  25 GeV and
  $\abs{\eta}<2.5$.
\item $\etmiss >25$ GeV.
\end{itemize}
These cuts accept 38\% of the $W_H\to Wh\to \ell\nu
b\overline{b}$ events. The missing transverse energy is assumed to
arise only from the neutrino in  the leptonic $W$  decay, and the
$W$ momentum is then reconstructed. Figure~\ref{wh2} shows  
 mass of the reconstructed $Wh$ system for $m_{W_H}=1000$ GeV and
 $\cot\theta=0.5$. The background which is dominated by $W+jets$ and
 $t\overline{t}$ events is
 larger than in the previous case, nevertheless a  clear signal is
 visible.

\begin{figure}
\centerline{\includegraphics[width=4in]{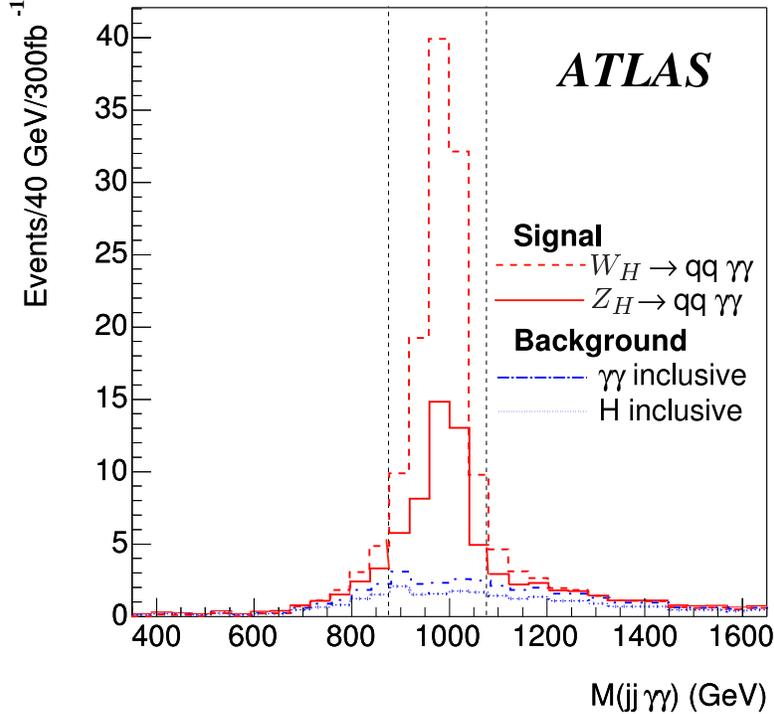}}
\caption{ Plot showing the invariant mass of $\gamma\gamma+Jets$
  system, where the  $\gamma\gamma$ system has a mass consistent with
  the Higgs mass of 120 GeV and the $Jet$ system is selected as
  described in the text. The largest histogram arises from $W_H\to h
  W$ and the second largest from $Z_H\to h
  Z$ for $\cot\theta= 0.5$ and a $Z_H$ and $W_H$ mass of 1000 GeV. The backgrounds from  Higgs
production (dotted) and the QCD production of di-photons (dashed) are
also shown. The vertical lines define the signal region.
 \label{azhgamgam}}
\end{figure}

The decay $h\to \gamma\gamma$ has a much smaller rate. Nevertheless it
provides a very characteristic signal. A preliminary event selection
requiring two isolated photons,  one having $p_T>25$ GeV and the other
$p_T>40$ GeV and both with  $\abs{\eta}<2.5$ was made. This requirement
ensures that the events are triggered. The invariant
mass of the two photon system is required to be within $2\sigma$ of the
Higgs mass, $\sigma$ being the measured mass resolution of the
diphoton system. The
reconstructed jets in the event  are then combined in pairs and the  pair with
invariant mass closest to $M_W$  was selected. If this pair has a
combined $p_T>200 $ GeV, its mass was corrected to the $W$ mass and then
combined with the $\gamma\gamma$ system. The mass distribution of the
resulting system is shown in Figure~\ref{azhgamgam} \cite{jose}. 
The $W$ from the
decay of $W_H$ is expected to have large $p_T$. Consequently it is
possible for  the jets from the $W$ have coalesced into a single
jet. To allow for this possibility, if the jet pair has
$p_T<200$ GeV it is not used.  Rather,  the  $\gamma\gamma$ system is combined with
any jet that has an invariant mass consistent with a $W$ and added to
Figure~\ref{azhgamgam}. This figure shows a reconstructed mass peak at the common mass of
$W_H$ and $Z_H$ of 1000 GeV; note that the states are expected to be
almost degenerate in the model. 50\% of the $\gamma\gamma+jet+jet$
signal events are accepted into the figure. The contribution from $W_H$ and $Z_H$ is
shown separately, the former dominates due to its larger production
rate. But, the mass resolution in the jet system is not good enough to
resolve the $W$ from the $Z$ when both decay hadronically. We cannot, therefore,  separate the
 signals from the $W_H$ and $Z_H$.
The presence of the two photons with a mass comparable to the
Higgs mass ensures that the background is small. The background
arises from either direct Higgs
production or the QCD production of di-photons, these are both shown
on the figure. The requirements on the jet system and photon acceptance
remove 98\% of the Higgs background. 

The signal to background ratio is large enough that a signal can also
be seen without reconstructing the jet system. Since the $W_H$ and $Z_H$ are
produced singly, with small transverse momentum, there is a
``Jacobian'' peak in the transverse momentum distribution of the
Higgs. This momentum is reconstructed with great precision from its
$\gamma\gamma$ decay mode owing to the excellent electromagnetic
calorimetry of the ATLAS detector. Of course, this inclusive signal
cannot distinguish $W_H$ from $Z_H$.  The inclusive nature of this
signal results in a small improvement in the efficiency 
as the jets from $W$ or $Z$ are not reconstructed. The sensitivity of this inclusive
signature is approximately the same as that from reconstructing the
hadronic decay of $W/Z$ \cite{jose}.

\begin{figure}
\centerline{\includegraphics[width=4in]{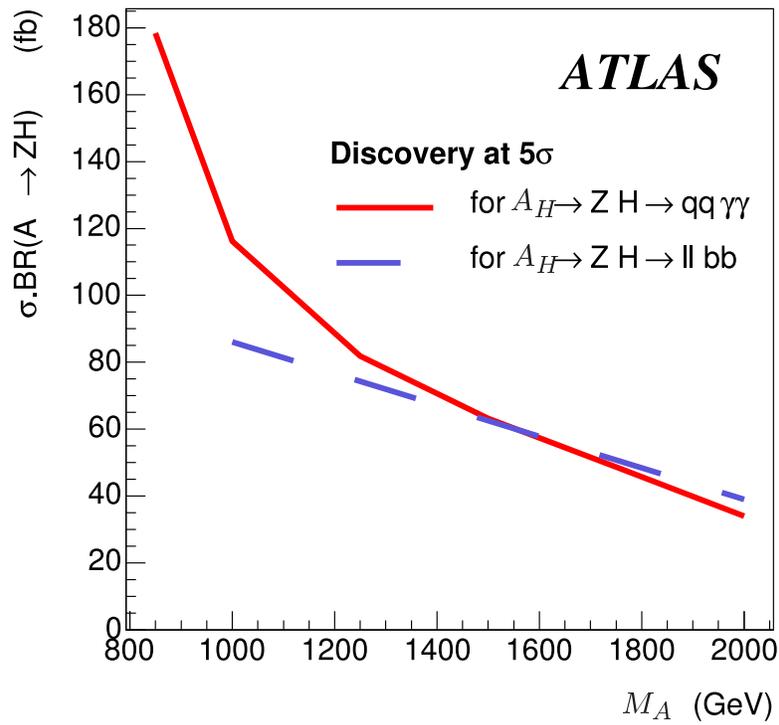}}
\caption{Plot showing the minimum  value of the production
cross-section times branching ratio needed to obtain discovery in the
channels $A_H\to Zh \to \ell\ell b\overline{b}$ and $A_H\to Zh \to
jets \gamma\gamma$ as a function of the $A_H$ mass.
 \label{ahlimit}}
\end{figure}

In the case of $A_H$ production and decay to $Zh$, the rates depend on
the mixings and so we present the sensitivity in terms of a
cross-section that allows  reinterpretation of these results to other
models.  Using the method described above, we show in Figure~\ref{ahlimit} the value of the production
cross-section times branching ratio needed to obtain discovery in the
channels $A_H\to Zh \to \ell\ell b\overline{b}$ and $A_H\to Zh \to
\hbox{jets } \gamma\gamma$.

\section{Search for $\phi^{++}$}

\begin{figure}
\centerline{\includegraphics[width=4in]{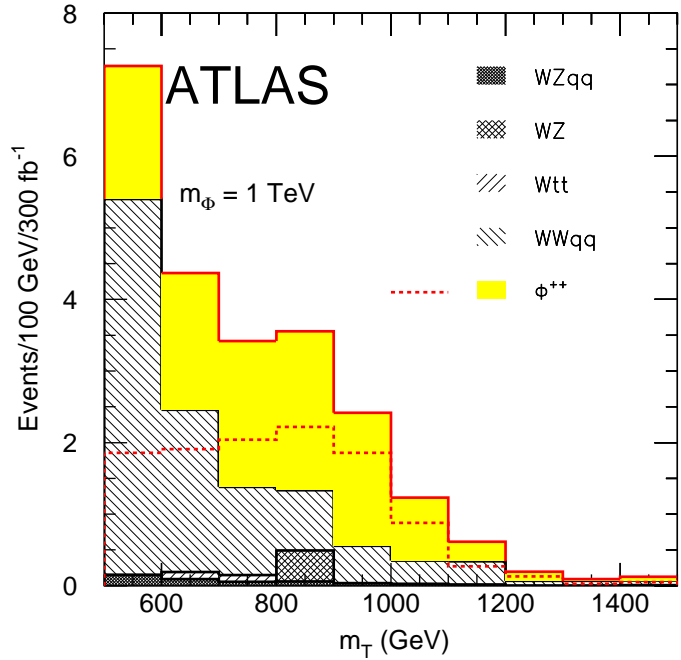}}
\caption{The mass distribution $M_T$, see text, for a $\phi^{++}$ of
  mass 1000 GeV and $v^{\prime}=25$ GeV. The dashed histogram shows 
the signal alone and the
  solid shows the sum of signal and backgrounds. The components of the
  background are also shown separately.
 \label{phipp1}}
\end{figure}

\begin{figure}
\centerline{\includegraphics[width=4in]{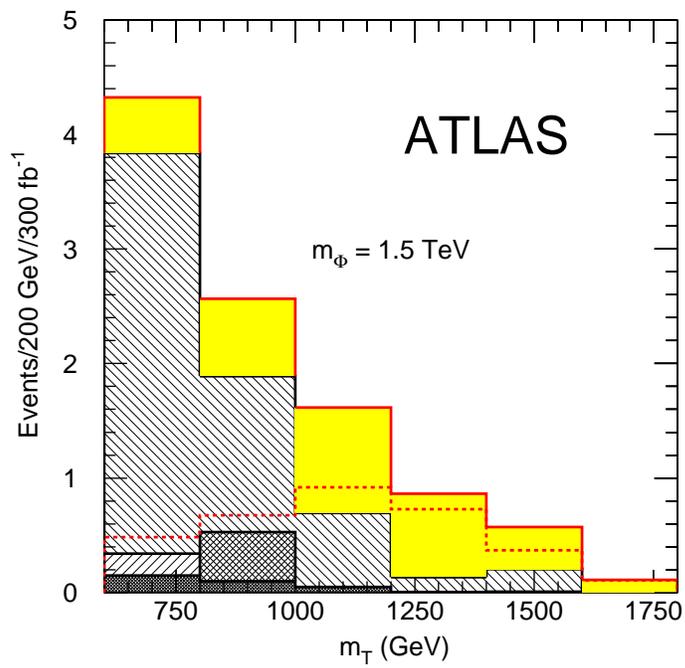}}
\caption{As in Figure~\ref{phipp1} except that the mass is 1500
  GeV. The legend is the same as on Figure~\ref{phipp1}.
 \label{phipp2}}
\end{figure}

The doubly-charged
Higgs boson could be  produced in pairs and decay into leptonic
final states via $
q \overline{q} \to \phi^{++}\phi^{--} \to 4 \ell$. While this  would provide a
very clean signature, it will not be considered here since the mass
reach in this channel  is poor due to the small cross-section.
The coupling of $\phi^{++}$ to $W^+W^+$ allows it to be produced singly
via
 $WW$ fusion processes of the type $dd\to uu \phi^{++} \to uu
W^+W^+$. This  can lead to events containing  two leptons of the same
charge, and missing energy from the
 decays of the $W$'s. The
$\phi WW$ coupling is determined by $v^{\prime}$, the  vacuum expectation value (vev)
of the
 neutral member of the triplet. This cannot be too large as its
 presence causes a violation of custodial SU(2) which is constrained by
 measurements of the $W$ and $Z$ masses. We have examined the
 sensitivity of searches at the LHC in terms of $v^{\prime}$ and the
 mass of $\phi^{++}$.  For $v^{\prime}= 25$ MeV and a mass of 1000
 TeV, the rate for production of $\phi^{++}$ followed by the decay to
 $WW$ is 4.9 fb if the $W$'s have $\abs{\eta} <3$ and $p_T>200 $ GeV. \cite{Han:2003wu}.
The simulation discussed below is normalized to this rate. As in the case of Standard Model Higgs searches using
 the $WW$ fusion process \cite{tdr}, the presence of jets
 at large rapidity must be used to  suppress backgrounds. The event
 selection closely follows that used in searches for a heavy Standard
 Model Higgs via the $WW$ fusion process \cite{wwfusion} and is  as follows.
 \begin{itemize}
 \item Two reconstructed positively charged isolated leptons (electrons or
   muons) with $\abs{\eta}<2.5$.
\item One of the leptons was required to have $p_T>150$  GeV and the
  other $p_T>20 $ GeV.
\item  The leptons are not balanced in transverse momentum: $\abs{p_{T1} - p_{T2}} > 200$ GeV.
\item The difference in pseudorapidity of the two leptons
  $\abs{\eta_1-\eta_2} < 2$.
\item $\etmiss >50 $ GeV.
\item Two jets each with $p_T>15$ GeV, with rapidities of opposite
  sign,  separated in rapidity
  $\abs{\eta_1-\eta_2} >5$; one jet has $E>200 $ GeV and the other
  $E>100$ GeV.
 \end{itemize}
The presence of the leptons ensures that the events are triggered. 
The
invarient mass of the $WW$ system cannot be reconstructed.
The signal can be observed  using a mass variable made from the observed
leptons  (${\bf p_1}$ and ${\bf p_2}$)
and the missing transverse energy as follows.
$$m^2_T = (E_1 + E_2 + \abs{\etmiss})^2 - ({\bf p_1}  + {\bf p_2} +
{\bf \etmiss })^2$$
The reconstructed mass distributions are shown in Figures~\ref{phipp1}
and \ref{phipp2} for masses of 1000 and 1500 GeV. Standard Model
backgrounds are shown separately on the figures \cite{kanaya}.  
In the region where $m^2_T>500$ GeV, the above kinematic cuts accept
50\% of the $\phi^{++}\to \ell^+ \nu\ell^+ \nu$ events for $\phi^{++}$
of mass 1 TeV.

The dominant background is from non-resonant $qqW^+W^+$ production, including
WW scattering as well as QED and QCD processes involving quark
scattering where $W$'s are
radiated from the quarks. The background  from the non-resonant $WW$
scattering, which  has the same topology
as the signal, was generated using CompHep \cite{Pukhov:1999gg}. 
In the region where $m^2_T>500$ GeV, the above kinematic cuts accept
5\% of the $WW\to \ell^+ \nu\ell^+ \nu$ process; the leptons being less
energetic that those from the signal. For a Higgs mass of 120 GeV, the contribution of
longitudinal WW scattering is very small. No $t\overline{t}$ events
survive the cuts; the forward jet tag being particularly effective in
this case \cite{kanaya}. The residual background from $WZ$ events is
below 10\% of that from  $qqW^+W^+$.
  Note that the rates shown in these figures are small
and the signal does not appear as a clear peak. The process is very
demanding of luminosity, the ability to detect forward jets at
relatively small $p_T$,  and the ability to control
backgrounds. These issues cannot be fully addressed until actual data
is available. At this stage, we can only estimate our sensitivity using
our current, best estimates, of the these issues. Since the cross-section
for a $\phi^{++}$ of a fixed mass is proportional to $(v^{\prime})^2$,
the simulation can be used to determine the sensitivity. Requiring at
least 10 events for $M_T> 700 (1000) $ GeV for $m_{\phi}=1000 (1500)$
GeV and a value of $S/\sqrt{B} > 5$  implies that discovery is
possible if $v^{\prime}> 29 (54)$ GeV.
Such values are larger than the constraint of 15 MeV from electro-weak
fits\cite{Han:2003wu}.

In view of this rather poor sensitivity, it is worth
considering other possible signals for the Higgs sector. The
$\phi^{+}$ and  $\phi^{0}$ can also be produced by gauge boson fusion
leading to $WZ$ and $W^+W^-$ final states. The latter is
similar qualitatively to the production of a heavy Standard Model
Higgs boson with reduced couplings. The background in these channels
is larger than that for $\phi^{++}$, so observation is likely to be
more difficult. Doubly charged Higgs bosons also occur in some
left-right symmetric models \cite{Huitu:1996su}. In these models the
boson can decay via a lepton flavor violating coupling to
$\ell^+\ell^+$. 
In the Littlest Higgs model under consideration 
the decay $\phi^{++}\to \ell^+\ell^+$ does not occur so this search is
not effective.

\section{Model Constraints and Conclusions}

We have demonstrated, using a series of examples,  how measurements using the ATLAS detector at the
LHC can be used to reveal the three characteristic particles of the Little
Higgs models. 
The $T$ quark is observable up to masses of
approximately 2.5 TeV via its decay to $Wb$. Sensitivity in   $Zt$ or
$Zh$ is lower but it still extends over the range expected in the
model provided that the Higgs mass is not too large. 
The decays of $T$ to $Wb$ and $Zt$ are, of course, independent of
the Higgs mass.  In the case of $Zh$ the sensitivity will depend on
the Higgs mass. The $h\to b\overline{b}$ channel continues to be
effective until Higgs mass exceeds 150 GeV or so when it becomes too
small to exploit due to the falling branching ratio for this
decay. The studies presented in section~\ref{sec:ht} are valid up to
this mass.  In this case ATLAS will be able to detect $T$ in its
three decay channels and provide a definitive test of the model.
For larger Higgs masses, the channels with excellent signal
to background ratios such
as $h\to \ell^+\ell^-\ell^+\ell^-$ and $h\to \ell^+\ell^-\nu\nu$, have
small effective branching ratios which will severely restrict a
search using them. However, ATLAS will still be able to detect the
$Wb$ and $Zt$ final states which will enable us to distinguish $T$
from a 4$^{th}$ generation quark.

\begin{figure}
\centerline{\includegraphics[width=4in]{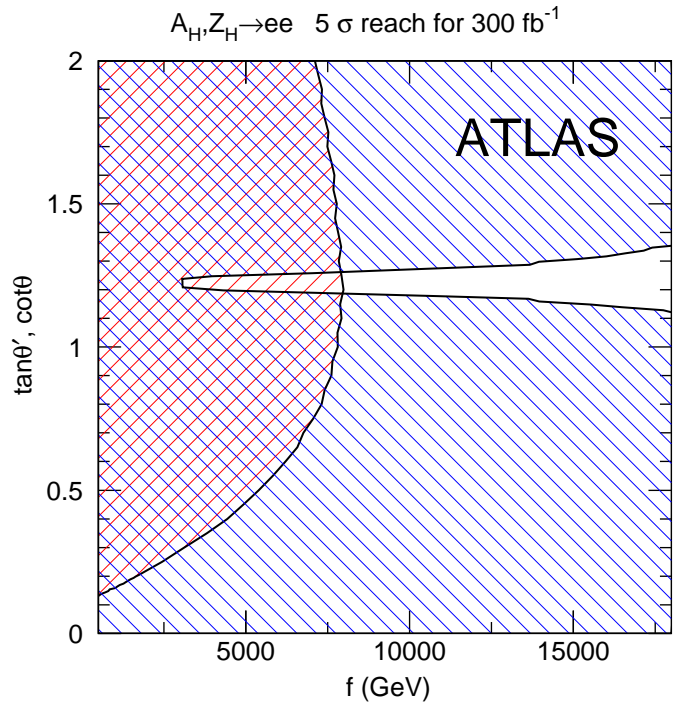}}
\caption{Plot showing the accessible regions for   $5\sigma$ discovery of the
  neutral gauge bosons $A_H$ and $Z_H$ as a function of the scale $f$
  where the additional symmetry is broken. The variable $\cot\theta$ 
  applies to the $Z_H$    for which the  red  region (lines
  running from bottom-left to top-right) is accessible. The blue  region (lines running
  from top-left to bottom-right) shows the $A_H$ accessible region as
  a function of $\tan\theta^\prime$.
 \label{reachf}}
\end{figure}

\begin{figure}
\centerline{\includegraphics[width=4in]{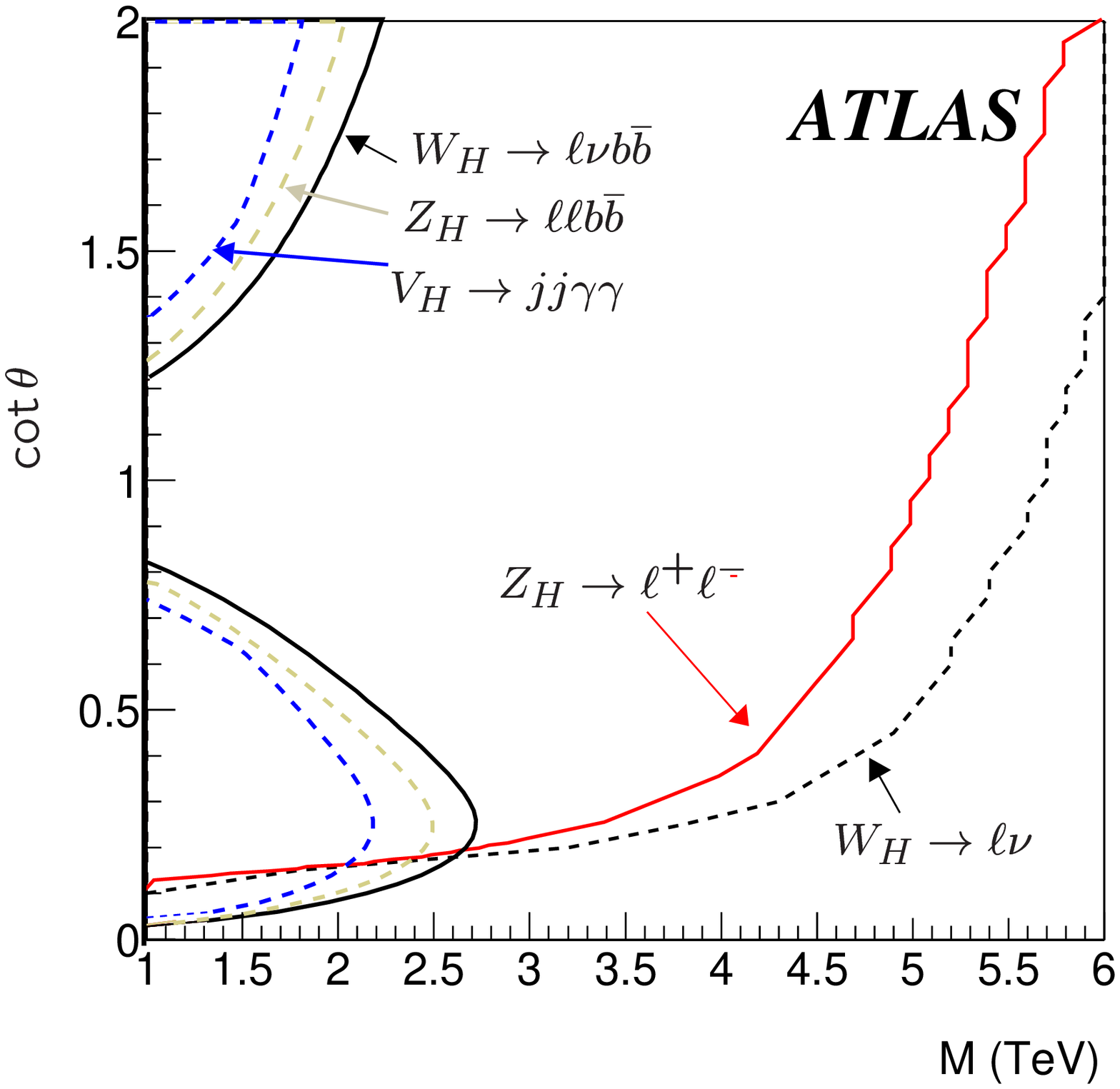}}
\caption{Plot showing the accessible  regions for $5\sigma$ discovery of the
  gauge bosons $W_H$ and $Z_H$ as a function of the mass and
  $\cot\theta$ for the various final states. The  regions to the left
  of the  lines are accessible with 300 $\fbi$.
 \label{reachwa}}
\end{figure}

In the case of the new gauge bosons, the situation is summarized in
Figures~\ref{reachf} and  \ref{reachwa}. The former shows the
accessible regions via the $e^+e^-$ final states of $Z_H$ and $A_H$ as
a function of the mixing angles and the scale $f$ that determines the
masses. Except for a small region near $\tan\theta^\prime = 1.3$ and
$f>7.5$ TeV, we are sensitive to the whole range expected in the model.
However observation of such a gauge boson will not prove that it is of
the type predicted in the Little Higgs Models. In order to do this, the
decays to the Standard Model bosons must be observed.
Figure~\ref{reachwa} shows the sensitive regions for decays of $Z_H$ and $W_H$
into various final states as a function of $\cot\theta$ and the
masses. 
It can be seen that several decay modes are only observable for
smaller masses over a restricted range of $\cot\theta$ where the characteristic
decays $Z_H\to Zh$ and $W_H\to Wh$ can be seen. There is a small
region  at very small values
of $\cot\theta$ where the leptonic decays are too small,  and only the
decays to $W/Z$ can be seen.  One of the channels shown on this plot,
the decay to $h$, is sensitive to the Higgs mass. The channel used for
this analysis, $h\to \gamma\gamma$ becomes less effective as the Higgs
mass increases. At the assumed mass of 120 GeV, the branching ratio is
approximately 2\%. As the rates shown in Figure~\ref{azhgamgam} scale
with this branching ratio, we expect that this mode is not visible for
Higgs masses above 150 GeV. Other decays of the Higgs have not yet
been investigated. To summarize, the new gauge bosons are observable
over the entire mass range predicted. Measurements of more than one
final state are possible in certain regions of parameter space.

In the case of $\phi^{++}$ the situation is not so promising. The
Higgs sector is the least 
constrained by fine tuning
arguments and  this
particle's mass can extend up to 10 TeV. We are only
sensitive to masses up to 2 TeV or so provided that $v^\prime$ is
large enough. Other ``Little Higgs'' models \cite{Kaplan:2003uc} have
a different Higgs structure that is similar to models with more than
one Higgs doublet that have been studied in the past. Work is needed
to evaluate the sensitivity of the LHC to these models.

In conclusion, we have investigated the signatures of the Littlest
Higgs model using ATLAS  at the LHC. We have seen that the LHC, with
its full luminosity, is fully sensitive to the new quarks and gauge
bosons. The new Higgs particles may be out of reach.

\section*{Acknowledgments}

This work has been performed within the ATLAS Collaboration, and we
thank collaboration members for helpful discussions.
We have made use of the physics analysis framework and tools which are
the result of collaboration-wide efforts.

This work was supported in part by the Director, Office
  of Science, Office of Basic Energy Sciences, of the U.S. Department
  of Energy under  Contract No. DE-AC03-76SF00098.
 Accordingly, the U.S.
Government retains a nonexclusive, royalty-free license to publish or
reproduce the published form of this contribution, or allow others to
do so, for U.S. Government purposes.

\end{document}